\documentclass[aps,
prd,twocolumn,floatfix,nofootinbib,showpacs,superscriptaddress,tightenlines]{revtex4}

\usepackage{amsmath}
\usepackage{lipsum}
\usepackage{amssymb}
\usepackage{amsthm}
\usepackage{bbold}
\usepackage{dcolumn}
\usepackage{epsfig}
\usepackage{graphics}
\usepackage{graphicx}
\usepackage{longtable}
\usepackage{color}

\usepackage{xspace}
\usepackage{cancel}

\definecolor{darkgreen}{rgb}{0,0.5,0}
\definecolor{purple}{rgb}{0.5,0,0.5}
\definecolor{nblue}{rgb}{0.0,0.0,0.50}
\definecolor{scarlet}{rgb}{1.0,0.2,0}

\usepackage[colorlinks=true, pdfstartview=FitV, linkcolor=purple, citecolor= purple, urlcolor=blue]{hyperref}

\newcommand{\me}{\mathrm{e}} %math e
\newcommand{\mi}{\mathrm{i}} %math i

\newcommand{\GeV}{\text{GeV}} %math GeV

\newcommand{\ie}{\text{i.e.}\xspace}
\newcommand{\eg}{\text{e.g.}\xspace}

\newcommand{\eqn}[1]{Eq.~(\ref{#1})}
\newcommand{\fig}[1]{Fig.~\ref{#1}}
\newcommand{\tab}[1]{Table~\ref{#1}}
\newcommand{\sect}[1]{Section~\ref{#1}}
\newcommand{\app}[1]{Appendix~\ref{#1}}

\newcommand{\eqnss}[2]{Eqs.~(\ref{#1},\ref{#2})}

\newcommand{\eqnsr}[2]{Eqs.~(\ref{#1}-\ref{#2})}

\newcommand{\df}[1]{\hspace{-0.5em}\ensuremath{\frac{\mathrm{d}^{4}#1}{(2\pi)^{4}}}\,}
\newcommand{\dx}[1]{\hspace{-0.5em}\ensuremath{\mathrm{d}#1}\,}

\newcommand{\Tr}{\text{Tr}}

\begin{document}

\title{Charmonia in a Contact Interaction}

\author{Marco A. Bedolla}
\author{J.J. Cobos-Mart\'{\i}nez}
\author{Adnan Bashir}
\affiliation{
Instituto de F\'{i}sica y Matem\'aticas, Universidad Michoacana de San
Nicol\'as Hidalgo, Apartados Postals 2-82, Morelia, Michoac\'an 58040, Mexico}

\date{\today}

\begin{abstract}
For the flavour-singlet heavy quark system of charmonia, we
compute the masses of the ground state mesons in four different
channels: pseudo-scalar ($\eta_c(1S)$), vector ($J/\Psi(1S)$),
scalar ($\chi_{c_0}(1P)$) and axial vector ($\chi_{c_{1}}(1P)$),
as well as the weak decay constants of the $\eta_c(1S)$ and
$J/\Psi(1S)$ and the charge radius of $\eta_c(1S)$. The framework
for this analysis is provided by a symmetry-preserving
Schwinger-Dyson equation (SDEs) treatment of a
vector$\times$vector contact interaction (CI). The results found
for the meson masses and the weak decay constants, for the
spin-spin combinations studied, are in fairly good agreement with
experimental data and earlier model calculations based upon
Schwinger-Dyson and Bethe-Salpeter equations (BSEs) involving
sophisticated interaction kernels. The charge radius of
$\eta_c(1S)$ is consistent with the results from refined SDE
studies and lattice Quantum Chromodynamics (QCD).
\end{abstract}


\maketitle

\date{\today}

\section{Introduction}

Quantum Chromodynamics (QCD) and the resulting hadron bound states
form a challenging sector of the Standard Model of particle
physics. In the non perturbative regime of these interactions, the
emergent phenomena of chiral symmetry breaking and confinement
govern their spectrum and properties. Within the framework of
Schwinger-Dyson equations (SDEs), we can study the structure of
strongly interacting bound states through first principles in the
continuum. SDEs for QCD have been extensively applied to the study
of light quark~\cite{Jain:1993qh,Maris:1997hd,Maris:1999nt} and
gluon
propagators~\cite{Boucaud:2008ky,Aguilar:2008xm,Pennington:2011xs},
their
interactions~\cite{Chang:2009zb,Kizilersu:2009kg,Bashir:2011dp},
meson spectra below the masses of 1 GeV as well as their static
and dynamic properties.

First explorations for heavy mesons, both charmonia and
bottomonia, with a consistent use of the rainbow-ladder truncation
in the kernels of the gap and Bethe-Salpeter equations (BSEs),
were undertaken by Jain and Munczek in Ref.~\cite{Jain:1993qh}.
They found the mass spectrum and the decay constants of
pseudoscalar mesons in good agreement with experiments. This work
was repeated with the Maris-Tandy model for $\bar{c}c$ bound
states in Refs.~\cite{Krassnigg:2004if, Bhagwat:2006xi} with
extrapolations to the $\bar{b}b$ systems in
Refs.~\cite{Maris:2005tt,Blank:2011ha}. A full numerical solution
for flavour-singlet pseudoscalar mesons again yielded charmonia
and bottomonia masses and decay constants consistent with
experimental data; predictions for states with exotic quantum
numbers were also made in Refs.~\cite{Maris:2006ea,
Krassnigg:2009zh}. The effect of the quark-gluon interaction in
the gap equation and the vertex-consistent Bethe-Salpeter kernel
was investigated in Ref.~\cite{Bhagwat:2004hn}. More recently,
employing a parametrization of the quark propagator to
analytically continue it into the complex plane, heavy
quark systems were studied in detail in
Ref.~\cite{Souchlas:2010zz}. A more direct approach through the
numerical computation of the quark propagator in the required
region of the complex plane was employed in Ref.~\cite{Rojas:2014aka}.
There the mass spectrum and decay constants for flavor singlet
$J^P=0^-$ mesons were reported.

The extension of this program to the complicated exotic and
baryonic states, decay rates and form factors is, numerically, not
straightforward at all. A few years ago, an alternative was
explored to initially study pion properties assuming that quarks
interact not via massless vector-boson exchange but instead
through a symmetry preserving vector-vector contact interaction
(CI)~\cite{GutierrezGuerrero:2010md,Roberts:2010rn,
Chen:2012qr,Roberts:2011cf,Roberts:2011wy}. One then proceeds by
embedding this interaction in a rainbow-ladder truncation of the
SDEs. Confinement is implemented by employing a proper time
regularization scheme. This scheme systematically removes
quadratic and logarithmic divergences ensuring that the
axial-vector Ward-Takahashi identity (axWTI) is satisfied. One can
also explicitly verify the low energy Goldberger-Treiman
relations. A fully consistent treatment of the CI model is simple
to implement and can help us provide useful results which can be
compared and contrasted with full QCD calculation and experiment.

This interaction is capable of providing a good description of the
masses of meson and baryon ground and excited-states for light
quarks~\cite{GutierrezGuerrero:2010md,Roberts:2010rn,
Chen:2012qr,Roberts:2011cf}. The results obtained from the CI
model are quantitatively comparable to those obtained using
sophisticated QCD model
interactions,~\cite{Bashir:2012fs,Eichmann:2008ae,Maris:2006ea,Cloet:2007pi}.
Interestingly and importantly, this simple CI produces a
parity-partner for each ground-state that is always more massive
than its first radial excitation so that, in the nucleon channel,
e.g., the first $J^P = {1}/{2}^-$ state lies above the second $J^P
= {1}/{2}^+$ state ~\cite{Chen:2012qr}.

We take this as a sufficient justification to employ this
interaction for the analysis of the quark model heavy mesons for
spins $J=0,1$ and study the mass spectrum and weak decay constants
for charmonia. Without parameter readjustment, we find good
agreement with charmonia masses. However, we need to modify the
set of parameters to simultaneously account for the weak decay
constants of the $\eta_c(1S)$ and $J/\Psi(1S)$, and the charge
radius of $\eta_c(1S)$.

The paper is organized as follows: in~\sect{sec:bse} we present
the necessary details of the SDE-BSE approach to mesons; while
in~\sect{sec:CI} and~\sect{sec:axwti} we introduce the interaction
used and the consequences of the axWTI; ~\sect{sec:bsaclas}
outlines the general forms of the Bethe-Salpeter amplitude (BSA)
for the mesons studied; \sect{sec:results} contains numerical
analysis of the results obtained without readjusting the
parameters of the light quark sector; while in
~\sect{sec:CINewModel} we minimally modify the CI model parameters
and re-evaluate the charmonia masses, decay constants of
$\eta_c(1S)$ and $J/\Psi(1S)$, and the charge radius of
$\eta_c(1S)$; finally in~\sect{sec:conclusions}, we state our
findings and present our conclusions and discussion.

\section{\label{sec:bse} The Bethe-Salpeter and the Gap Equations}

Meson bound states appear as poles in a four-point function. The
condition for the appearance of these poles in a particular
$J^{PC}$ channel is given by the homogeneous BSE~\cite{Gross:1993zj,
Salpeter:1951sz,GellMann:1951rw}
 \begin{equation}
 \label{eqn:bse}
 \left[\Gamma_{H}(k;P)\right]_{tu}=
 \int\df{q}\chi(q;P)_{sr}K^{rs}_{tu}(q,k;P) \,,
 \end{equation}
where $\chi(q;P)=S_{f}(q_{+})\Gamma_{H}(q;P)S_{g}(q_{-})$;
$q_{+}=q+\eta P$, $q_{-}=q-(1-\eta)P$; $k$ ($P$) is the relative
(total) momentum of the quark-antiquark system; $S_{f}$ is the
$f$-flavour quark propagator; $\Gamma_{H}(q;P)$ is the meson BSA,
where $H=f\overline{g}$ specifies the flavour content of the
meson; $r,s,t,u$ represent colour, flavour, and spinor indices;
and $K$ is the quark-antiquark scattering kernel. For a
comprehensive recent review of the BSE and its applications,
see~\cite{Bashir:2012fs}.

The $f$-flavour dressed-quark propagator, $S_{f}$, that enters
\eqn{eqn:bse} is obtained as the solution of the quark SDE, the so
called gap
equation~\cite{Roberts:2007jh,Holl:2006ni,Maris:2003vk,Alkofer:2000wg}:
 \begin{eqnarray}
\label{eqn:quark_sde}
&& \hspace{-0.5cm} S_{f}^{-1}(p)=i\gamma\cdot p + m_{f} + \Sigma_{f}(p) \,, \\
\label{eqn:quark_se} && \hspace{-0.3cm}
\Sigma_{f}(p)=\int\df{q}g^{2}D_{\mu\nu}(p-q)\frac{\lambda^{a}}{2}
\gamma_{\mu}S_{f}(q)\Gamma^{a}_{\nu}(p,q) \,,
 \end{eqnarray}
where $g$ is the strong coupling constant, $D_{\mu\nu}$ is the
dressed-gluon propagator, $\Gamma^{a}_{\nu}$ is the
dressed-quark-gluon vertex, and $m_{f}$ is the bare $f$-flavour
current-quark mass. Since the CI, to be defined later in
\sect{sec:CI}, is non-renormalizable, it is not necessary to
introduce any renormalziation constant, and the chiral limit is
obtained by setting
$m_{f}=0$~\cite{Roberts:2007jh,Holl:2006ni,Maris:2003vk}.

Both $D_{\mu\nu}$ and $\Gamma^{a}_{\nu}$ satisfy their own SDE,
which in turn are coupled to higher $n$-point functions and so on
{\it ad infinitum}. Therefore, the quark SDE, \eqn{eqn:quark_sde},
is only one of the infinite set of coupled nonlinear integral
equations. A tractable problem is defined once a truncation scheme
has been specified, \ie, once the gluon propagator and the
quark-gluon vertex are defined.

\section{\label{sec:CI} Rainbow-Ladder truncation and the Contact Interaction}

%Lite sector mass spectrum----------------------------------------------------
\begin{table}[h]
\begin{center}
\begin{tabular}{lllllllll}
\hline
\hline
$m$ & $M$ & $E_{\pi}$ & $F_{\pi}$ & $E_{\rho}$ & $m_{\pi}$ & $m_{\rho}$
& $f_{\pi}$ & $f_{\rho}$ \\
\hline
0 & 0.358 & 3.568 & 0.459 & 1.520 & 0 & 0.919 & 0.100 & 0.130 \\
0.007 & 0.368 & 3.639 & 0.481 & 1.531 & 0.140 & 0.928 & 0.101 & 0.129 \\
\hline
\hline
\end{tabular}
\caption{\label{tab:muu} Results for light pseudoscalar and vector
mesons obtained with $m_{g}=0.8\,\GeV$, $\alpha_{IR}=0.93\pi$,
$\Lambda_{\text{IR}}=0.24\,\GeV$,
$\Lambda_{\text{UV}}=0.905\,\GeV$. These model parameters were
determined in Refs.~\cite{Roberts:2011cf,Chen:2012qr}}
\end{center}
\end{table}
%----------------------------------------------------------------------------

It has been shown ~\cite{GutierrezGuerrero:2010md,Roberts:2010rn,
Chen:2012qr,Roberts:2011cf} that a momentum-independent
vector$\times$vector CI is capable of providing a description of
light pseudoscalar and vector mesons static properties, which is
comparable to that obtained using more sophisticated QCD model
interactions
\cite{Bashir:2012fs,Eichmann:2008ae,Maris:2006ea,Cloet:2007pi};
see for example \tab{tab:muu}. We employ this interaction for the
analysis of the quark model charmonia spectrum. We therefore use
 \begin{eqnarray}
\label{eqn:contact_interaction} g^2 D_{\mu \nu}(k) =
\frac{4\pi\alpha_{\rm IR}}{m_g^2}\delta_{\mu \nu} \equiv
\frac{1}{m_{G}^{2}}\delta_{\mu\nu} \,,
 \end{eqnarray}
in \eqn{eqn:quark_se}, where $m_g=800$ MeV is a gluon mass scale
which is in fact generated dynamically in QCD, see for
example~\cite{Boucaud:2011ug}, and $\alpha_{\rm IR}=0.93 \pi$  is
a parameter that determines the interaction strength. For the
quark-gluon vertex, the rainbow truncation will be used:
 \begin{equation}
\label{eqn:quark_gluon_vertex_rl}
\Gamma^{a}_{\mu}(p,q)=\frac{\lambda^{a}}{2}\gamma_{\mu} \,.
 \end{equation}
Once the elements of the kernel in the quark SDE have been
specified, we can proceed to obtain and analyse its solution. The
general form of the f-flavoured dressed quark propagator, obtained
as the solution of \eqn{eqn:quark_sde}, is given in terms of two
Lorentz-scalar dressing functions, written in two different but
equivalent forms as:
 \begin{eqnarray}
\label{eqn:gen_quark_inverse}
S^{-1}_{f}(p)&=& \mi\gamma\cdot pA_{f}(p^{2}) + B_{f}(p^{2}) \\
        &=& Z^{-1}_{f}(p^{2})\left(i\gamma\cdot p + M_{f}(p^{2})
        \right) \, .
 \end{eqnarray}
In the latter expression, $Z(p^{2})$ is known as the wave-function
renormalization, and $M_{f}(p^{2})$ is the dressed,
momentum-dependent quark mass function, which connects current and
constituent quark
masses~\cite{Roberts:2007jh,Holl:2006ni,Maris:2003vk}.

Using \eqn{eqn:contact_interaction} and \eqn{eqn:quark_gluon_vertex_rl}, the
quark equation, \eqn{eqn:quark_sde}, takes the following simple form
 \begin{equation}
\label{eqn:quark_sde_contact} S^{-1}_{f}(p)=i\gamma\cdot p + m_{f}
+
\frac{4}{3}\frac{1}{m_{G}^{2}}\int\df{q}\gamma_{\mu}S_{f}(q)\gamma_{\mu}
\,.
 \end{equation}
The solution to this equation is of the form
 \begin{equation}
\label{eqn:quark_inverse_contact} S_{f}^{-1}(p)= \mi\gamma\cdot p
+ M_{f} \,,
 \end{equation}
\noindent since the last term on the right-hand side of
\eqn{eqn:quark_sde_contact} is independent of the external
momentum. The momentum-independent mass, $M_{f}$, is determined as
the solution of
\begin{equation}
\label{eqn:const_mass} M_{f} = m_{f} +
\frac{M_{f}}{3\pi^{2}m_{G}^{2}}\int_{0}^{\infty}\dx{s}s\frac{1}{s+M_{f}^{2}}
\,.
\end{equation}
Since \eqn{eqn:const_mass} is divergent, we have to specify a
regularization procedure. We employ the proper time regularization
scheme,~\cite{Ebert:1996vx}, and write:
\begin{eqnarray}
\label{eqn:regularisation_procedure}
\frac{1}{s+M^{2}}&=&\int_{0}^{\infty}\dx{\tau}\me^{-\tau(s+M^{2})}
\to
\int_{\tau_{\rm{UV}}^{2}}^{\tau_{{\rm{IR}}^{2}}}\dx{\tau}\me^{-\tau(s+M^{2})}, \nonumber \\
&=& \frac{\me^{-\tau_{\text UV}^2(s+M^{2})}-\me^{-\tau_{\text
IR}^2(s+M^{2})}}{s + M^2} \,,
\end{eqnarray}
where $\tau_{\text{IR}}^{2}$ and $\tau_{\text{UV}}^{2}$ are,
respectively, infrared and ultraviolet regulators. A nonzero value
for $\tau_{\text{IR}}\equiv 1/\Lambda_{\text{IR}}$ implements
confinement by ensuring the absence of quark production thresholds
~\cite{Roberts:2007ji}. Furthermore, since
\eqn{eqn:contact_interaction} does not define a renormalizable
theory, $\tau_{\text{UV}}\equiv 1/\Lambda_{\text{UV}}$ cannot be
removed, but instead plays a dynamical role and sets the scale for
all dimensioned quantities. Note that the role of ultraviolet
cut-off in Nambu--Jona-Lasinio type models has also been discussed
in Refs.~\cite{Farias:2005cr,Farias:2006cs}. Thus
\begin{equation}
\label{eqn:const_mass_reg} M_{f}= m_{f} +
\frac{M_{f}}{3\pi^{2}m_{G}^{2}}\mathcal{C}_{01}(M_{f}^{2};\tau_{\text{IR}},
\tau_{\text{UV}}) \,,
\end{equation}
where
\begin{equation}
\label{eqn:Cfun}
\mathcal{C}_{\alpha\beta}(M^{2};\tau_{\text{IR}},\tau_{\text{UV}})=
\frac{\left(M^{2}\right)^{\nu}}{\Gamma(\beta)}
\Gamma(\beta-2,\tau_{\text{UV}}^{2}M^{2},\tau_{\text{IR}}^{2}M^{2})
\,,
\end{equation}
with $\nu=\alpha-(\beta-2)$ and  $\Gamma(a,z_{1},z_{2})$ is the
generalized incomplete gamma function.

\section{\label{sec:axwti} Axial-Vector Ward-Takahashi Identtiy}

The phenomenological features of chiral symmetry and its dynamical
breaking in QCD can be understood  by means of the axWTI. In the
chiral limit, it reads
 \begin{equation}
\label{eqn:axwti} -\mi P_{\mu}\Gamma_{5\mu}(k;P) =
S^{-1}(k_{+})\gamma_{5} + \gamma_{5}S^{-1}(k_{-}) \,.
 \end{equation}
The axWTI relates the axial-vector vertex, $\Gamma_{5\mu}$, the
pseudoscalar vertex, $\gamma_{5}$ and the quark propagator. This
in turn implies a relationship between the kernel in the BSE and
that in the quark SDE. It must be preserved by any viable
truncation scheme of the SDE-BSE coupled system. It is the
preservation of this identity which proves useful in obtaining the
defining characteristics of the octet of pseudoscalar mesons, namely
their low mass, their masslessness in the chiral limit, and the hadron mass
splittings~\cite{Maris:1997hd,Weise:2005kf}

The axial-vector vertex satisfies its own SDE, namely
 \begin{equation}
\label{eqn:psvsde}
\Gamma_{5\mu}(k;P)= \gamma_{5}\gamma_{\mu}
                 + \int\df{q}K(k,q;P)\chi_{5\mu}(q;P) \,,
\end{equation}
where the appropriate indices are contracted and $K(k,q;P)$ is the
Bethe-Salpeter kernel that appears in the bound state
\eqn{eqn:bse}. A similar equation is satisfied by the pseudoscalar
vertex.

Combining the SDEs satisfied by the pseudovector and pseudoscalar
vertices with the axWTI one arrives at~\cite{Maris:1997hd}
 \begin{multline}
\label{eqn:axialwti_kernels_rels}
\int\df{q}K_{tu;rs}(k,q;P)\left[\gamma_{5}S(q_{-})
+ S(q_{+})\gamma_{5}\right]_{sr} \\
= \left[\Sigma(k_{+})\gamma_{5} + \gamma_{5}\Sigma(k_{-})
\right]_{tu} \, ,
\end{multline}
thus constraining  the content of the quark-antiquark scattering
kernel $K(p,q;P)$ if an essential symmetry of the strong
interactions, and its breaking pattern, is to be faithfully
reproduced.

From a practical point of view, \eqn{eqn:axialwti_kernels_rels}
provides a way of obtaining the quark-antiquark scattering kernel
if we can solve this constraint, given an expression for the quark
self-energy. However, this is not always possible, see \eg
~\cite{Fischer:2007ze}, and we must find an alternative way of
preserving the chiral symmetry properties of the strong
interactions. In principle, one may construct a quark-antiquark
scattering kernel satisfying \eqn{eqn:axialwti_kernels_rels} from
a functional derivative of the quark self-energy with respect to
the quark propagator~\cite{Munczek:1994zz}, within the framework
of the effective action formalism for composite operators
developed in~\cite{Cornwall:1974vz}.

Fortunately, for the CI model under study,
\eqn{eqn:axialwti_kernels_rels} can be easily satisfied. The
resulting expression for the scattering kernel is called the
rainbow-ladder (RL) truncation. This kernel is the leading-order term
in a non perturbative, symmetry-preserving truncation scheme,
which is known and understood to  be accurate for pseudoscalar and
vector mesons. Moreover, it guarantees electromagnetic current
conservation~\cite{Roberts:2007ji}:
 \begin{equation}
\label{eqn:bskernel_rl_contact} K(p,q;P)_{tu;rs}= -
\frac{1}{m_{G}^{2}}\delta_{\mu\nu}
\left[\frac{\lambda^{a}}{2}\gamma_{\mu}\right]_{ts}
\left[\frac{\lambda^{a}}{2}\gamma_{\nu}\right]_{ru} \,.
\end{equation}
Using the interaction we have specified,
\eqnss{eqn:contact_interaction}{eqn:quark_gluon_vertex_rl}, the
homogeneous BSE for a meson ($\eta=1$) takes a simpler form,
\begin{equation}
\label{eqn:bse_contact} \Gamma_{H}(k;P)=
-\frac{4}{3}\frac{1}{m_{G}^{2}}\int\df{q}
\gamma_{\mu}S_{f}(q+P)\Gamma_{H}(q;P)S_{g}(q)\gamma_{\mu} \,.
\end{equation}
\noindent Since the interaction does not depend on the relative
momentum of the quarks, a symmetry-preserving regularization of
\eqn{eqn:bse_contact} will yield solutions which are independent
of it. It follows that if the interaction of \eqn{eqn:contact_interaction}
produces bound states, then the relative momentum between the quark and the
antiquark can assume any value with equal probability. This is the defining
characteristic of a point-like particle.

\subsection{A Corollary of the Axial-Vector WTI}

There are further non trivial consequences of the axWTI and the
CI. They define our regularization procedure, which must maintain
 \begin{equation}
\label{eqn:wticorollary} 0 =\int\df{q} \left[\frac{P\cdot
q_{+}}{q_{+}^{2}+M_{f}^{2}} - \frac{P\cdot q}{q_{-}^{2}+M_{g}^{2}}
\right] \,.
 \end{equation}
This ensures that~\eqn{eqn:axwti} is satisfied. Now analyzing the
integrands, using a Feynman parametrization, one arrives at the
following identity for $P^{2}=0=m$, and $M_{f}=M_{g}=M$:
 \begin{equation}
\label{eqn:wticorollary2} 0 =\int\df{q} \frac{\frac{1}{2}q^{2} +
M^{2}}{\left(q^{2}+M^{2}\right)^{2}} \,.
 \end{equation}
\eqn{eqn:wticorollary2}  states that the axWTI is satisfied if,
and only if, the model is regularized so as to ensure there are no
quadratic or logarithmic divergences. Unsurprisingly, these are
the circumstances under which a shift in integration variables is
permitted, an operation required in order to prove
\eqn{eqn:axwti}.

It is notable that \eqn{eqn:axwti} is also valid for arbitrary
$P$. Using a Feynman parametrization of the integrand, and making
an appropriate change of variables ($q\to q-xP$) to diagonalize
the denominator, we find, for non zero $P^{2}$
 \begin{equation}
\label{eqn:wticorollary3} 0 =\int_{0}^{1}\dx{x}\int\df{q}
\frac{\frac{1}{2}q^{2} + \mathfrak{M}^{2}} {\left(q^{2}+
\mathfrak{M}^{2}\right)^{2}} \,,
 \end{equation}
where $\mathfrak{M}^{2}=M_{f}^{2}x + M_{g}^{2}(1-x)+ x(1-x)P^{2}$.
This constraint will be implemented in all our calculations so
that~\eqn{eqn:axwti} is preserved.

\section{\label{sec:bsaclas} Classification of BSA in a Contact Interaction}

We are interested in the static properties of several mesons. We
thus begin with their classification and the general form of their
BSA in the CI we are working with.
\tab{tab:mesons} lists the spin quantum numbers of the quark model
mesons we will study.

%----------------------------------------------------------------------------
\begin{table}[h]
\begin{center}
\begin{tabular}{llllll}
\hline
\hline
$L$ & $J^{PC}$ & Type & $L$ & $J^{PC}$ & Type \\
\hline
$0$ & $0^{-+}$ & Pseudoscalars & $1$ & $0^{++}$ & Scalars \\
$0$ & $1^{--}$ & Vectors       & $1$ & $1^{++},\,1^{+-}$ & Axial Vectors \\
\hline
\hline
\end{tabular}
\caption{\label{tab:mesons} Quark model mesons}
\end{center}
\end{table}
%---------------------------------------------------------------------------

With the dependence on the relative momentum forbidden by the CI,
the general form of the BSAs for the mesons listed in
\tab{tab:mesons} are ~\cite{LlewellynSmith:1969az}:
\begin{eqnarray}
\label{eqn:psbsagral}
\Gamma_{0^{-+}}(P)&=& \gamma_{5}\left[\mi E_{0^{-+}}
+ \frac{1}{2M}\gamma\cdot P F_{0^{-+}}\right] \,, \\
\label{eqn:sbsagral}
\Gamma_{0^{++}}(P)&=& \mathbb{1}E_{0^{++}} \,, \\
\label{eqn:vbsagral}
\Gamma_{1^{--}\mu}(P)&=&\gamma^{T}_{\mu}E_{1^{--}}
+ \frac{1}{2M}\sigma_{\mu\nu}P_{\nu}F_{1^{--}} \,, \\
\label{eqn:avbsagral}
\Gamma_{1^{++}\mu}(P)&=&\gamma_{5}\left[\gamma^{T}_{\mu}E_{1^{++}}
+ \frac{1}{2M}\sigma_{\mu\nu}P_{\nu}F_{1^{++}}\right] \,,
\end{eqnarray}
\noindent where $M$ is a mass scale, to be defined later. Results
will be independent of its choice. A charge-conjugated BSA is
obtained, in general, via
\begin{equation}
\label{eqn:chargecojugation}
\overline{\Gamma}_{H}(k;P)=C^{\dagger}\Gamma(-k;P)^{T}C,
\end{equation}
\noindent where $T$ denotes the transposing of all matrix indices and
$C=\gamma_{2}\gamma_{4}$ is the charge conjugation matrix, with
$C^{\dagger}=-C$, and $[C,\gamma_{5}]= 0$. We thus have
\begin{eqnarray}
\label{eqn:ccids}
C^{\dagger}\gamma_{\mu}^{T}C &=&-\gamma_{\mu} \,,  \\
C^{\dagger}\sigma_{\mu\nu}^{T}C &=&-\sigma_{\mu\nu} \,,  \\
C^{\dagger}\gamma_{5}^{T}C &=&\gamma_{5} \,, \\
C^{\dagger}\gamma_{5}\sigma^{T}_{\mu\nu}C &=&-\gamma_{5}\sigma_{\mu\nu} \,, \\
C^{\dagger}(\gamma_{5}\sigma_{\mu\nu})^{T}C
&=&\gamma_{5}\gamma_{\mu} \,,
\end{eqnarray}
\noindent and therefore
\begin{eqnarray}
\label{eqn:ccbsa}
\overline{\Gamma}^{0^{++}}(P)&=& \Gamma^{0^{++}}(P), \\
\overline{\Gamma}^{1^{--}}_{\mu}(P)&=& -\Gamma^{1^{--}}_{\mu}(P), \\
\overline{\Gamma}^{1^{++}}_{\mu}(P)&=& \Gamma^{1^{++}}_{\mu}(P).
\end{eqnarray}

It is a well known feature of the rainbow-ladder
truncation of the SDE-BSE system that it describes the
pseudoscalar and vector mesons well, but not their parity
partners, namely, the scalar and axial-vector mesons. In more
realistic kernels, see for example~\cite{Chang:2009zb}, when the
quark-gluon vertex is fully dressed, it was found that dynamical
chiral symmetry breaking (DCSB) generates a large dressed-quark
anomalous chromo-magnetic moment in the infrared. Consequently,
the associated corrections cancel in the pseudoscalar and vector
channels but add in the scalar and axial vector channels,
resulting in a magnified splitting between parity partners. This
effect is specially important for mesons made up of light quarks.
With this in mind, following Ref.~\cite{Roberts:2011cf}, we have
introduced a spin-orbit repulsion into the scalar- and
axial-vector-meson channels through the artifice of a
phenomenological coupling $g_{SO}^{2}\le 1$, introduced as a
factor multiplying the scalar and axial-vector kernels. The value
$g_{SO}=0.24$ was chosen in Ref.~\cite{Roberts:2011cf} so as to
obtain the experimental value for the $a_{1}$-$\rho$ mass
splitting which is known to be achieved by the corrections
described above (without the spin-orbit coupling $g_{SO}$, the
mass difference between the $a_{1}$ and the $\rho$ is 0.15 GeV, a
factor of 3 smaller than the experimental value, It implies that
the spin-orbit coupling has increased the mass of the $a_1$ by
$28\%$, which is a large effect).\normalcolor

\section{\label{sec:results} Numerical Results}

The mass and BSA of a meson depend on its quantum numbers and can
be found by solving \eqn{eqn:bse}. In order to do this, we will
introduce a fictitious eigenvalue $\lambda_{H}$ to the bound state
equation. Thus, the mass of the bound state in a particular
channel, $m_{H}$, will be such that $\lambda_{H}(P^{2}=-m_{H}^{2})=1$, where
$P$ is the meson's momentum. In any channel, the form of the homogeneous BSE
for the CI will be
 \begin{equation}
\label{eqn:generic_bse}
K_{H}(m_{H})\cdot\Gamma_{H}(m_{H})=\lambda_{H}(m_{H})\Gamma_{H}(m_{H})
\,,
 \end{equation}
where $K_{H}$ is a $2\times 2$ matrix, and  the subscript $H$
indicates the dependence of the explicit expressions on the
quantum numbers of the meson under consideration, see
\eqnsr{eqn:psbsagral}{eqn:avbsagral}. Equation
(\ref{eqn:generic_bse}) is an eigenvalue equation for the vector
$\Gamma_{H}(m_{H})=(E_{H}(m_{H}),F_{H}(m_{H}))^{T}$ with solutions
for discrete values of $P^{2}=-m_{H}^{2}$. Explicit expressions
for every channel given in \tab{tab:mesons} are presented in
\app{app:kernels}.

%Lite sector parameter set results---------------------------------------------
\begin{table}[h]
\begin{center}
\begin{tabular}{lllll}
\hline
\hline
    & \multicolumn{4}{c}{masses}  \\
\hline
& $m_{\eta_{c}(1S)}$ & $m_{J/\Psi(1S)}$ & $m_{\chi_{c_{0}}(1P)}$ & $m_{\chi_{c_{1}}(1P)}$
\\
\hline
Experiment ~\cite{0954-3899-37-7A-075021} & 2.983 & 3.096 & 3.414 & 3.510  \\
Contact Interaction & 2.983$^{*}$ & 2.979 & 3.412 & 3.442\\
                    & - & - & 3.293 & 3.344 \\
JM~\cite{Jain:1993qh} & 2.821 & 3.1 & 3.605 & - \\
BK~\cite{Blank:2011ha} & 2.928 & 3.111 & 3.321 & 3.437 \\
S1rp~\cite{Souchlas:2010zz} & 3.035 & 3.192 & - & - \\
RB1~\cite{Rojas:2014aka} & 3.065 & - & - & - \\
RB2~\cite{Rojas:2014aka} & 3.210 & - & - & - \\
\hline \hline \hline
\end{tabular}
\caption{\label{tab:mcc_all_lite} Ground state charmonia masses
obtained with the light sector parameter set: $m_{g}=0.8\,\GeV$,
$\alpha_{IR}=0.93\pi$, $\Lambda_{\text{IR}}=0.24\,\GeV$,
$\Lambda_{\text{UV}}=0.905\,\GeV$. The current-quark mass is
$m_{c}=1.578^{*}\,\GeV$, and the dynamically generated
constituent-like mass is $M_{c}=1.601\,\GeV$. The value
immediately below the CI results is obtained without a spin-orbit
coupling $g_{so}=0.24$. For a direct comparison, we quote values
from other SDE approaches to calculate the masses of low lying
charmonia. Dimensioned quantities are in GeV. ($^{*}=$ The current
quark mass was fitted to obtain the mass of the pseudoscalar
meson).}
\end{center}
\end{table}
%---------------------------------------------------------------------------

%Lite sector parameter set results--------------------------------------------
\begin{figure}[ht]
\includegraphics[width=0.53\textwidth]{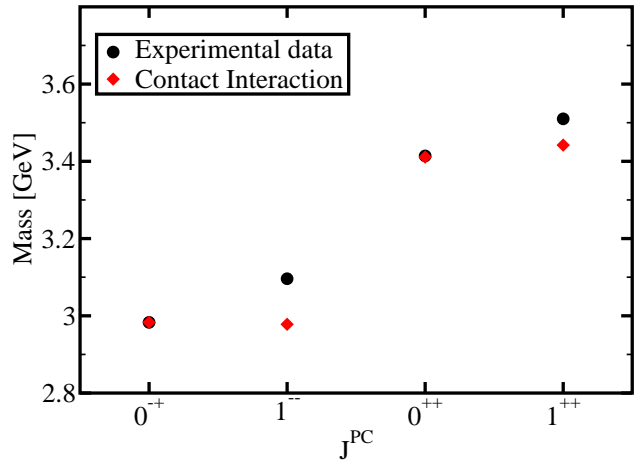}
\caption{\label{fig:cbarc_mass_spectrum}Contact interaction results for the
$\overline{c}c$ mass spectrum using model parameters fitted in the light
sector, see \tab{tab:mcc_all_lite}
PDG-labeled data is from ~\cite{0954-3899-37-7A-075021}.}
\end{figure}
%-----------------------------------------------------------------------------

\subsection{\label{sec:masses} Charmonia mass spectrum}

It has been shown~\cite{GutierrezGuerrero:2010md,Roberts:2010rn,
Chen:2012qr,Roberts:2011cf} that a momentum-independent
vector$\times$vector interaction is capable of providing a
description of light pseudoscalar and vector mesons static
properties which is comparable to that obtained using more
sophisticated QCD model interactions
~\cite{Bashir:2012fs,Eichmann:2008ae,Maris:2006ea,Cloet:2007pi}.
Here we assess the capability of this model to provide a
description of static properties of charmonia. The parameter set
used in this calculation is the same as that obtained used in
the light sector. Only the current-quark mass for the charm quark is
an input parameter, and it is fixed such that the experimental mass of the
pseudoscalar is reproduced. The rest of the meson masses are predictions of the
model. As can be seen from \tab{tab:mcc_all_lite}, the
predictions for the masses of the remaining mesons are in good
agreement with the results obtained from more sophisticated
SDE-BSE model calculations~\cite{Blank:2011ha,Rojas:2014aka}, lattice QCD
for the charm sector~\cite{Follana:2006rc,Wagner:2013laa} as well as
experimental values~\cite{0954-3899-37-7A-075021}.

That a RL truncation with a CI describes well the mass spectrum of
ground state charmonia can be understood in a simple way. Since
the wave function renormalization and quark mass function are
momentum-independent the heavy quark-gluon vertex can therefore
reasonably be approximated by a bare vertex, ensuring the vector
and axWTI.

\subsection{\label{sec:decays}Decay constants}

%Lite sector parameter set results--------------------------------------------
\begin{table}[h]
\begin{center}
\begin{tabular}{lllll}
\hline
\hline
    & \multicolumn{4}{c}{amplitudes}  \\
\hline
& $\eta_{c}(1S)$ & $J/\Psi(1S)$ & $\chi_{c_{0}}(1P)$ & $ \chi_{c_{1}}(1P)$
\\
\hline
%Experiment ~\cite{0954-3899-37-7A-075021} & 2.983 & 3.096 & 3.414 & 3.510  \\
%$m_{H}$ & 2.983$^{*}$ & 2.979 & 3.412 & 3.442 \\
%       & - & - & 3.293 & 3.344 \\
%\hline
$E_{H}$ & 6.028 & 3.024 & 0.437 & 0.298 \\
       & -  & - & 1.905 & 1.153 \\
$F_{H}$ & 1.711 & - & - & - \\
\hline
    & \multicolumn{4}{c}{decay constants}  \\
\hline
Experiment ~\cite{0954-3899-37-7A-075021} & 0.361 & 0.416 & - & -\\
Contact Interaction & 0.084 & 0.080 & - & - \\
\hline
\hline
\end{tabular}
\caption{\label{tab:Acc_all_lite}  Ground state charmonia
amplitudes and decay constants obtained with the light sector
parameter set adopted in \tab{tab:mcc_all_lite}:
%$m_{g}=0.8\,\GeV$, $\alpha_{IR}=0.93\pi$, $\Lambda_{\text{IR}}=0.24\,\GeV$,
%$\Lambda_{\text{UV}}=0.905\,\GeV$. The current-quark mass is
%$m_{c}=1.578^{*}\,\GeV$, and the dynamically generated constituent-like mass
%is $M_{c}=1.601\,\GeV$. The value immediately below the CI results is obtained
%without a spin-orbit coupling $g_{so}=0.24$.
%Dimensioned quantities are in GeV. ($^{*}=$ The current quark mass was fitted
%to obtain the mass of the pseudoscalar meson).
}
\end{center}
\end{table}
%--------------------------------------------------------------------------

Since the BSE is a homogeneous equation, the BSA has to be
normalized by a separate condition. In the Rainbow-Ladder
truncation of the BSE, that condition takes a simple form (we
choose $\eta=1$):
 \begin{equation}
\label{eqn:RLNorm} \hspace{.2cm}
P_{\mu}=N_{c}\frac{\partial}{\partial
P_{\mu}}\hspace{-1mm}\int\df{q} \hspace{-1mm}
\Tr\left[\overline{\Gamma}_{H}(-Q)S(q_{+})\Gamma_{H}(Q)S(q)\right]
\,,
 \end{equation}
at $Q=P$, with $P^{2}=-m_{H}^{2}$, which ensures that the residue
at the mass pole is unity. Here, $\Gamma_{H}$ is the normalized
BSA and $\overline{\Gamma}_{H}$ its charge conjugated version.

For every channel, we will re-scale $\Gamma_{H}$ such that
\eqn{eqn:RLNorm} is satisfied. Thus, we replace $\Gamma_{H}$ with
$\Gamma_{H}/N_{H}$, where $N_{H}$ is the normalization constant
and now $\Gamma_{H}$ is the non-normalized BSA, the amplitude that
is obtained by solving the homogeneous BSE. Thus the normalization
constant is obtained from
 \begin{equation}
\label{eqn:RLNormConst}
N_{H}^{2}P^{2}=N_{c}P_{\mu}\frac{\partial}{\partial
P_{\mu}}\hspace{-1mm} \int\df{q}\hspace{-1mm}
\Tr\left[\overline{\Gamma}_{H}(-Q)S(q_{+})\Gamma_{H}(Q)S(q)\right]
\,,
 \end{equation}
at $Q=P$, with $P^{2}=-m_{H}^{2}$. For the vector and axial vector
channels there is an additional factor of $1/3$ on the right hand
side since we have to take into account all three meson
polarizations.

Once the BSA has been normalized canonically, we can calculate
observables from it. The pseudoscalar leptonic decay constant,
$f_{0^{-+}}$, is defined by (in a more realistic interaction,
there is a factor of $Z_{2}$ on the right-hand side, and of course
the BSA depends on the relative momentum)
\begin{equation}
\label{eqn:psdecaydef}
P_{\mu}f_{0^{-+}}=N_{c}\int\df{q}\Tr\left[\gamma_{5}\gamma_{\mu}
S(q_{+})\Gamma_{0^{-+}}(P)S(q_{-})\right] \,.
\end{equation}
Similarly, the vector decay constant, $f_{1^{--}}$, is defined by
 \begin{equation}
\label{eqn:vdecaydef}
m_{1^{--}}f_{1^{--}}=\frac{N_{c}}{3}\int\df{q}\Tr\left[\gamma_{\mu}
S(q_{+})\Gamma_{\mu 1^{--}}S(q_{-})\right] \,,
 \end{equation}
where $m_{1^{--}}$ is the mass of the $1^{--}$ bound state, and
the factor of $3$ in the denominator comes from summing over the
three polarizations of the spin-1 meson. Explicit expressions for
the normalization condition in every channel and the decay
constants $f_{0^{-+}}$ and $f_{1^{--}}$ are given in
\app{app:norms}.

As can be seem from \tab{tab:Acc_all_lite}, the pseudoscalar and
vector decay constants, for the model parameters used, are
strongly underestimated, in disagreement both with model
calculations and experimental data. Numerically, this is because
the corresponding BSAs are too small. Changing
$\Lambda_{\text{UV}}$, e.g., to $1.843\,\GeV$, keeping the other
parameters fixed, except for the current-quark masses, which are
taken from Ref.~\cite{0954-3899-37-7A-075021}, improves the
situation by about a factor of $2$. However, there is still a
significant mismatch between our results, model calculations,
and experiment. We observe that this disagreement persists
despite the fact that our calculation for the pseudoscalars are in perfect
agreement with the Gell-Mann--Oakes--Renner relation, which is
valid for every $0^{-}$ meson, irrespective of the magnitude of
the current-quark mass,~\cite{Krassnigg:2009gd,Holl:2004fr,Blank:2011ha}.

It is not difficult to understand why the decay constants come out
to be much smaller than what one expects from the QCD based SDE
with running quark mass function. As noticed in
Refs.~\cite{Bhagwat:2004hn,Maris:1997tm,Maris:1999nt}, the decay
constant is influenced by the high momentum tails of the
dressed-quark propagator and the BSAs. This high momentum region
probes the wave-function of quarkonia at origin. The CI, on the
other hand, yields constant mass with no perturbative tail for
large momenta. Therefore, this artefact of quarkonia has to be
built into the model in an alternative manner. We know that with
increasing mass of the heavy quarks, they become increasingly
point-like in the configuration space. The closer the quarks get,
the further the coupling strength between them decreases.
Therefore, we cannot expect the decay constants to be correctly
reproduced with the parameters of the light quark sector. In the
next section, we consider the possibility of extending the simple
CI model to the heavy sector by reducing the effective coupling.
However, the reduction in the strength of the kernel has to be
compensated by increasing the ultraviolet cut-off. This makes
sense by observing that the $\Lambda_ {\text{UV}}$ (highest energy
scale associated with the system) used in the light quark sector
is, in fact, less than the current charm quark mass. Therefore, it
needs to be modified. We look for a  balance between the effective
coupling and the ultraviolet cut-off to describe the static
properties of charmonia.

\section{\label{sec:CINewModel} Contact Interaction Model for Charmonia}

As we mentioned in the last section, we set out to redefine the
parameters of the CI to study the masses, weak decay constants and
the charge radii of charmonia.

%\subsection{\label{sec:Charmonia} Charmonia}

%Optimal parameter set results----------------------------------------------
\begin{table}[h]
\begin{center}
\begin{tabular}{lllll}
\hline
\hline
    & \multicolumn{4}{c}{masses}  \\
\hline
& $m_{\eta_{c}(1S)}$ & $m_{J/\Psi(1S)}$ & $m_{\chi_{c_{0}}(1P)}$ & $m_{\chi_{c_{1}}(1P)}$
\\
\hline
Experiment ~\cite{0954-3899-37-7A-075021} & 2.983 & 3.096 & 3.414 & 3.510  \\
%Contact Interaction & 2.958$^{*}$ & 3.300 & 3.517 & 3.544 \\
%                    &  &  & 3.386 & 3.449 \\
Contact Interaction & 2.950$^{*}$ & 3.129 & 3.407 & 3.433 \\
                    &  &  & 3.194 & 3.254 \\
JM ~\cite{Jain:1993qh} & 2.821 & 3.1 & 3.605 & - \\
BK ~\cite{Blank:2011ha} & 2.928 & 3.111 & 3.321 & 3.437 \\
RB1 ~\cite{Rojas:2014aka} & 3.065 & - & - & - \\
RB2 ~\cite{Rojas:2014aka} & 3.210 & - & - & - \\
\hline \hline \hline
\end{tabular}
\caption{\label{tab:mcc_all_opt} Ground state charmonia masses
obtained with the best-fit parameter set: $m_{g}=0.8\,\GeV$,
$\alpha_{IR}=0.93\pi/20$, $\Lambda_{\text{IR}}=0.24\,\GeV$,
$\Lambda_{\text{UV}}=2.788\,\GeV$. The current-quark mass is
$m_{c}=0.956^{*}\,\GeV$, and the dynamically generated
constituent-like mass is $M_{c}=1.497\,\GeV$. The value
immediately below the CI results is obtained without a spin-orbit
coupling $g_{so}=0.24/3$. Dimensioned quantities are in GeV.
($^{*}=$ This parameter set was obtained from a best-fit to the
mass and decay constant of the pseudoscalar and vector channels).}
\end{center}
\end{table}
%---------------------------------------------------------------------------

%Optimal parameter set results-------------------------------------------------
\begin{figure}[ht]
\includegraphics[width=0.53\textwidth]{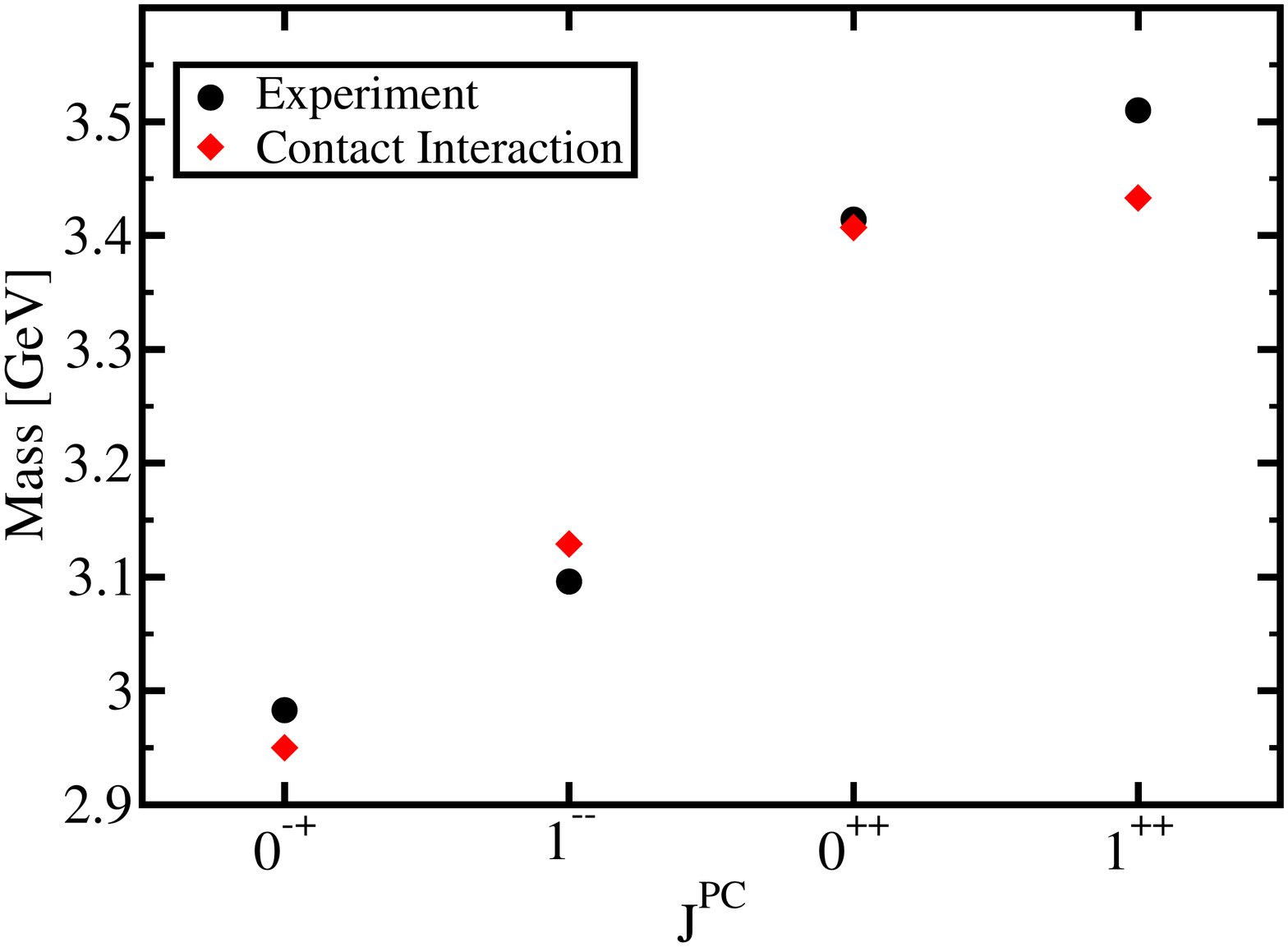}
\caption{\label{fig:cbarc_mass_spectrum_opt}Contact interaction
results for the $\overline{c}c$ mass spectrum, see
\tab{tab:mcc_all_opt}. PDG-labelled data is taken from
Ref.~\cite{0954-3899-37-7A-075021}.}
\end{figure}
%-----------------------------------------------------------------------------

%Optimal parameter set results-----------------------------------------------
%A table could be put here
%--------------------------------------------------------------------------

We retain the parameters $m_g$ and $\Lambda_{\text{IR}}$ of the
light sector. Modern studies of the gluon propagator indicate that
in the infrared, the dynamically generated gluon mass scale
virtually remains unaffected by the introduction of heavy
dynamical quark masses, see for example
Refs.~\cite{Ayala:2012pb,Bashir:2013zha}. The rest of the
parameters are obtained from a best-fit to the mass and decay
constant of the pseudoscalar ($\eta_{c}$) and vector ($J/\Psi$)
channels.

One can now readily calculate the masses of the ground state
pseudo-scalar, vector, scalar, and axial vector mesons. The
results are shown in \tab{tab:mcc_all_opt} and
\fig{fig:cbarc_mass_spectrum_opt}. They are in very good agreement
with experimental values and comparable to the best SDE results
with refined truncations. It is true that the masses of
charmonia from quenched quark models can be shifted by large
amounts when considering hadron loops, see e.g.
\cite{Barnes:2007xu}. This observation appears to invalidate the
quenched quark model results. However, as noted in
Ref.~\cite{Barnes:2007xu}, since this scale of mass shift is
common to all low-lying states, it can therefore be absorbed in a
change of model parameters. Thus, instead of consistently adding
hadron loops into our Contact Interaction calculation, we have
mimicked the effect by fitting the model parameters (coupling
constant and ultraviolet cutoff). It ensures, for example, a
constituent-like charm quark mass of the order of 1 GeV and a
correct value for the experimental mass of the
$\eta_{c}$.\normalcolor

For the case of the $\chi_{c_{1}}$, we find a mass of
3.254 GeV without a spin-orbit coupling, which is $7\%$ lower than
the experimental value; our calculated value is even closer to
that of the experimental $J/\psi$ mass. A similar pattern is
observed with the pseudoscalar and scalar channels. Therefore, to
achieve an acceptable mass difference between parity partners, we
have introduced a spin-orbit coupling of $g_{SO}=0.08$. This gives
the values in the second row of~\tab{tab:mcc_all_opt}. As can be
seen from the these values, the mass of the $\chi_{c_{1}}$ has
increased only by $5\%$. This small effect is in line with the
heavy-quark spin symmetry results. The spin-dependent interactions
are proportional to the chromomagnetic moment of the quark
(simulated by a spin orbit coupling). Predictably, these effects
are substantially suppressed for charmonia as compared to light
mesons. \normalcolor

The decay constants for the $\eta_{c}(1S)$ and
$J/\Psi(1S)$ channels are reported in Table~\ref{tab:DecayCNew}.
For the pseudoscalar meson, the result aligns nicely with the experimental
value. However, this is not exactly the case for the vector channel.
Furthermore, we note that the decay constant for $J/\Psi(1S)$ is smaller than
that for $\eta_{c}(1S)$. The correct ordering can be recovered by reducing
the interaction strength by a large factor. However, this is
something we consider contrived and, therefore, not pursue
further. Notice that one of the SDE results yields the
$J/\Psi(1S)$ decay constant even smaller than our
value,~\cite{Souchlas:2010zz}.

%-----------------------------------------------------------------------------
\begin{table}[h]
\begin{center}
\begin{tabular}{lllll}
\hline \hline
    & \multicolumn{3}{c}{decay constants}  \\
\hline \hline
 & $f_{\eta_{c}}$ & $f_{J/\Psi}$
\\
Experiment ~\cite{0954-3899-37-7A-075021} & 0.361 & 0.416 \\
S1rp ~\cite{Souchlas:2010zz} & 0.239 & 0.198 \\
S3ccp ~\cite{Souchlas:2010zz} & 0.326 & 0.330 \\
BK ~\cite{Blank:2011ha} & 0.399 & 0.448 \\
Contact Interaction & 0.305 & 0.220  \\
\hline \hline
\end{tabular}
\caption{\label{tab:DecayCNew} The decay constants for the states
$\eta_{c}(1S)$ and $J/\Psi(1S)$ obtained with $m_{g}=0.8\,\GeV$,
$\alpha_{IR}=0.93\pi/20$, $\Lambda_{\text{IR}}=0.24\,\GeV$,
$\Lambda_{\text{UV}}=2.788\,\GeV$.
The current-quark mass is $m_{c}=0.956\,\GeV$. Dimensioned quantities are in
GeV.}
\end{center}
\end{table}
%---------------------------------------------------------------------------

%-----------------------------------------------------------------------------
\begin{table}[h]
\begin{center}
\begin{tabular}{lllll}
\hline \hline
   & \multicolumn{3}{c} {$\eta_{c}(1S)$ charge radius } \\
\hline \hline
 & SDE~\cite{Bhagwat:2006xi} & Lattice ~\cite{Dudek:2007zz} & CI  \\
 &  0.219 {\text fm} & 0.25 {\text fm} & 0.21 {\text fm} \\
 \hline \hline
\end{tabular}
\caption{\label{tab:ChargeradiusC} The charge radius for the state
$\eta_{c}(1S)$ obtained with the new parameter set discussed in
the text.}
\end{center}
\end{table}
%---------------------------------------------------------------------------

As another test for the new parameter set for the CI Model for
charmonia, we compute the charge radius of the $\eta_{c}(1S)$,
$r^{2}_{\eta_{c}}=-6\partial F_{\eta_{c}}(Q^{2})/\partial Q^{2}|_{Q^{2}=0}$ where
$F_{\eta_{c}}(Q^{2})$ is its electromagnetic form factor which we will report
elsewhere and compare it with the results presented in
Ref.~\cite{Bhagwat:2006xi} from previous SDE studies and
Refs.~\cite{Bhagwat:2006xi,Dudek:2007zz} from Lattice QCD. The results are
presented in \tab{tab:ChargeradiusC}. As can be seen, our calculated
charge radius is very close to the one obtained by employing the Maris-Tandy
Model and the one reported in the lattice
studies~\cite{Bhagwat:2006xi,Dudek:2007zz}.

\newpage

\section{\label{sec:conclusions} Conclusions}

We compute the quark model ground state spin-0 and spin-1
charmonia  masses and decay constants using a rainbow-ladder
truncation of the simultaneous set of SDE and BSE with a CI model
of QCD, developed and tested for the light quark
sector~\cite{GutierrezGuerrero:2010md,Roberts:2010rn,
Chen:2012qr,Roberts:2011cf,Roberts:2011wy}. As the model is
non-renormalizable, we employ proper time regularization scheme
which ensures confinement is implemented through the absence of
quark production threshold. Moreover, the relevant Ward identities
and the low energy theorems such as Goldberger-Triemann relations
are satisfied. Without parameter readjustment, we find that the
masses  of the studied mesons are in reasonably good agreement
with experimental data and other model calculations. Moreover, the
Gell-Mann--Oakes--Renner relation, valid for every current-quark
mass in the pseudo-scalar channel, is always satisfied. However,
the decay constants of pseudo-scalar as well as vector mesons are
significantly underestimated.

We realize that the extension of the CI model to the heavy sector
requires a reduction of the effective coupling, which mimics the
high momentum tail of the quark mass function obtained in the SDE
studies of QCD~\cite{Bhagwat:2004hn,Maris:1997tm,Maris:1999nt}. We
only have to ensure that the reduction in the strength of the
kernel is appropriately compensated by increasing the ultraviolet
cut-off, a natural requirement for studying heavy quarks. We find
that with a modified choice of two parameters, not only the masses
of the ground state mesons, i.e., pseudo-scalar ($\eta_c(1S)$),
vector ($J/\Psi(1S)$), scalar ($\chi_{c_0}(1P)$), and axial vector
($\chi_{c_{1}}(1P)$), but also their weak decay constants, are in
much better agreement with the experiments~\cite{0954-3899-37-7A-075021}
as well as earlier SDE calculations with QCD based refined
truncations~\cite{Blank:2011ha}. As a further test of the model,
we evaluate the charge radius of $\eta_c(1S)$ and found it
reasonably close to the results computed in
Refs.~\cite{Bhagwat:2006xi,Dudek:2007zz}. This is an encouraging
first step towards a comprehensive study of heavy mesons in this
approach. Further steps will involve flavored mesons and baryons.
Our goal is to provide a unified description of light and heavy
hadrons within the CI model.

\appendix

\section{\label{app:kernels} BSE Kernels}

Here we give explicit expressions for the kernel in every channel
considered in this article. The general expression for the kernel
of matrix elements is

\begin{widetext}
\begin{eqnarray}
\label{eqn:Hkernelgral}
\mathcal{K}^{ij}_{H}&=&-\frac{4}{3}\frac{1}{m_{G}^{2}}
\int\df{q}
\frac{\Tr\left[\mathcal{P}^{i}_{H}(P)\gamma_{\mu}\hat{S}_{f}(q_{+})
\mathcal{D}^{j}_{H}(P)\hat{S}_{g}(q_{-})\gamma_{\mu}\right]}
{(q_{+}^{2}+M_{f}^{2})(q^{2}+M_{g}^{2})} \\
\label{eqn:Hkernelgralpars} &=&-\frac{4}{3}\frac{1}{m_{G}^{2}}
\int_{0}^{1}\dx{x}\int\df{q}
\frac{\Tr\left[\mathcal{P}^{i}_{H}(P)\gamma_{\mu}\hat{S}_{f}(q_{+})
\mathcal{D}^{j}_{H}(P)\hat{S}_{g}(q_{-})\gamma_{\mu}\right]_{q\to
q-xP}} {(q^{2}+\mathfrak{M}^{2})^{2}} \,, \nonumber
\end{eqnarray}
\end{widetext}

\noindent where $\hat{S}_{f}(k)=-\mi\gamma\cdot k + M_{f}$, and
$\mathcal{D}_{H}^{i}$, $\mathcal{P}_{H}^{i}$ are, respectively,
suitable Dirac covariants projectors for a given channel. In order
to write compact expressions below, we define the following
expressions

\begin{align}
J(x)&=x(1-x)\,,   \hspace{-1cm} &K(x)&=M_{f}x+M_{g}(1-x)\,, \nonumber \\
L(x)&=M_{f}^{2}x+M_{g}^{2}(1-x)\,, &L^{\pm}(x)&=M_{f}M_{g}\pm L(x)
\,.
 \nonumber
\end{align}
For notational simplicity, we shall omit an overall factor of
${1}/{3\pi^{2}m_{G}^{2}}$ which multiplies the kernel in every
channel.

\subsection{Pseudoscalar kernel}

 For the pseudoscalar channel,
\begin{align}
\label{eqn:psdirac12} \mathcal{D}^{1}_{0^{-+}}&=\mi\gamma_{5} \,,
&\mathcal{D}^{2}_{0^{-+}}&= \frac{1}{2M}\gamma_{5}\gamma\cdot P \,,\\
\label{eqn:psprojF12}
 \mathcal{P}^{1}_{0^{-+}}&=-\frac{\mi}{4}\gamma_{5} \,,
& \mathcal{P}^{2}_{0^{-+}}&=-\frac{M}{2P^{2}}\gamma_{5}\gamma\cdot
P \,.
\end{align}
\noindent Thus

\vspace{-5mm}

\begin{eqnarray}
\label{eqn:psK11} && \hspace{-0.4cm} \mathcal{K}^{11}_{0^{-+}}=
\hspace{-0.1cm} \int_{0}^{1}\dx{x} \hspace{-0.1cm}
\left[\mathcal{C}_{01}(\mathfrak{M}^{2}) \hspace{-0.1cm} +
\hspace{-0.1cm} \left(L^{-}(x)-
2J(x)P^{2}\right)\mathcal{C}_{02}(\mathfrak{M}^{2})\right] ,
\nonumber \\
\\
\label{eqn:psK12} && \hspace{-0.4cm}
\mathcal{K}^{12}_{0^{-+}}=\frac{P^{2}}{2M}
\int_{0}^{1}\dx{x}K(x)\mathcal{C}_{02}(\mathfrak{M}^{2}) \,,
\\
\label{eqn:psK21} && \hspace{-0.4cm}  \mathcal{K}^{21}_{0^{-+}}=M
\int_{0}^{1}\dx{x}K(x)\mathcal{C}_{02}(\mathfrak{M}^{2}) \,,
\\
\label{eqn:psK22} && \hspace{-0.4cm}
 \mathcal{K}^{22}_{0^{-+}}=-\frac{1}{2}
\int_{0}^{1}\dx{x}L^{+}(x)\mathcal{C}_{02}(\mathfrak{M}^{2}) \,.
\end{eqnarray}

\subsection{Vector kernel}

For the vector channel,
\begin{align}
\label{eqn:vdirac12} \mathcal{D}^{1}_{1^{--}}&=\gamma^{T}_{\mu}
\,,
&\mathcal{D}^{2}_{1^{--}}&=\frac{1}{2M}\sigma_{\mu\nu}P_{\nu} \,,\\
\label{eqn:vprojF12}
 \mathcal{P}^{1}_{1^{--}}&=\frac{1}{12}\gamma^{T}_{\mu} \,,
& \mathcal{P}^{2}_{1^{--}}&=\frac{M}{6P^{2}}\sigma_{\mu\nu}P_{\nu}
\,.
\end{align}
Thus
\begin{eqnarray}
\label{eqn:vK11} && \hspace{-0.6cm}
\mathcal{K}^{11}_{1^{--}}=\frac{1}{2}
\int_{0}^{1}\dx{x}(L^{-}(x)-2J(x)P^{2})\mathcal{C}_{02}(\mathfrak{M}^{2})
\,,
\\
\label{eqn:vK12} && \hspace{-0.6cm}
\mathcal{K}^{12}_{1^{--}}=-\frac{P^{2}}{4M}
\int_{0}^{1}\dx{x}K(x)\mathcal{C}_{02}(\mathfrak{M}^{2}) \,,
\\
\label{eqn:vK21} && \hspace{-0.6cm} \mathcal{K}^{21}_{1^{--}}= 0
\,,
\\
\label{eqn:vK22} && \hspace{-0.6cm} \mathcal{K}^{22}_{1^{--}}= 0
\,.
\end{eqnarray}

\subsection{Scalar kernel}

For the scalar channel,
\begin{eqnarray}
\label{eqn:sdirac1}
\mathcal{D}^{1}_{0^{++}}&=& \mathbb{1} \,, \\
\label{eqn:sprojF1}
\mathcal{P}^{1}_{0^{++}}&=&\frac{1}{4}\mathbb{1} \,.
\end{eqnarray}
Thus
\begin{eqnarray}
\label{eqn:sK11} \mathcal{K}^{11}_{0^{++}}= \hspace{-0.2cm}
\int_{0}^{1}\dx{x}\left[\mathcal{C}(\mathfrak{M}^{2})-
(L^{+}(x)+2J(x)P^{2})\mathcal{C}_{02}(\mathfrak{M}^{2})\right] \,.
\nonumber \\
\end{eqnarray}

\subsection{Axial vector kernel}

For the axial vector channel,
\begin{align}
\label{eqn:avdirac12}
\mathcal{D}^{1}_{1^{++}}&=\gamma_{5}\gamma^{T}_{\mu} \,,
&\mathcal{D}^{2}_{1^{++}}&=\gamma_{5}\frac{1}{2M}\sigma_{\mu\nu}P_{\nu} \,,\\
\label{eqn:avprojF12}
 \mathcal{P}^{1}_{1^{++}}&=\frac{1}{12}\gamma^{T}_{\mu}\gamma_{5} \,,
&
\mathcal{P}^{2}_{1^{++}}&=\frac{M}{6P^{2}}\sigma_{\mu\nu}P_{\nu}\gamma_{5}
\,.
\end{align}
Thus
\begin{eqnarray}
\label{eqn:avK11} \mathcal{K}^{11}_{1^{++}}&=&-\frac{1}{2}
\int_{0}^{1}\dx{x}(L^{+}(x)+2J(x)P^{2})\mathcal{C}_{02}(\mathfrak{M}^{2})
\,,
\nonumber \\
\\
\label{eqn:avK12} \mathcal{K}^{12}_{1^{++}}&=& 0 \,,
\\
\label{eqn:avK21} \mathcal{K}^{21}_{1^{++}}&=& 0 \,,
\\
\label{eqn:avK22} \mathcal{K}^{22}_{1^{++}}&=& 0 \,.
\end{eqnarray}

\section{\label{app:norms} Normalization}

In the appendix, we give explicit expressions for normalization
condition in every channel considered and for the decay constants
of pseudoscalar and vector mesons. The general expression for the
normalization condition can be written as
\begin{equation}
\label{eqn:norm_cond_gral}
N_{H}^{2}=\sum\limits_{i,j=1}\mathcal{N}^{ij}_{H}
\mathcal{F}^{i}_{H}\mathcal{F}^{j}_{H} \,,
\end{equation}
\noindent where the upper limits on the summations depend on the
number of non zero dressing functions in a given channel. A factor
of $1/16\pi^2$ multiplies every $\mathcal{N}^{ij}_{H}$ and recall
the factor of $1/3$ for the vector and axial vector channels,
stemming from all three polarizations.

\subsection{Pseudoscalar channel}

\noindent In this channel, the normalizations are:
\begin{align}
\label{eqn:psnormconst11} \hspace{-0.5cm}
\mathcal{N}^{11}_{0^{-+}}= ~ ~ ~  8 ~ ~ ~\int_{0}^{1}\dx{x}J(x)&
\left[2\left(L^{-}(x)-2J(x)P^{2}\right)\mathcal{C}_{03}(\mathfrak{M}^{2})
\right.  \nonumber \\
& \left.+3\mathcal{C}_{02}(\mathfrak{M}^{2})\right] \,, \\
\label{eqn:psnormconst12}
\mathcal{N}^{12}_{0^{-+}}=-\frac{2}{M}\int_{0}^{1}\dx{x}K(x)
&\left[-4J(x)P^{2}\mathcal{C}_{03}(\mathfrak{M}^{2})\right. \nonumber \\
&\left.+ \mathcal{C}_{02}(\mathfrak{M}^{2})\right] \,, \\
\label{eqn:psnormconst21}
\mathcal{N}^{21}_{0^{-+}}=-\frac{2}{M}\int_{0}^{1}\dx{x}K(x)
&\left[-4J(x)P^{2}\mathcal{C}_{03}(\mathfrak{M}^{2})\right. \nonumber \\
&\left.+ \mathcal{C}_{02}(\mathfrak{M}^{2})\right] \,,\\
\label{eqn:psnormconst22}
\mathcal{N}^{22}_{0^{-+}}=-\frac{4P^{2}}{M^{2}}
\int_{0}^{1}\dx{x}J(x)&L^{+}(x)\mathcal{C}_{03}(\mathfrak{M}^{2})
\,.
\end{align}

\subsubsection{Pseudoscalar decay constant}

\noindent Explicit expression for the decay constant is as
follows:
\begin{eqnarray}
\label{eqn:psdecayexplfinal}
f_{0^{-}}&=& N_{c}\left[\mathcal{K}_{0^{-}}^{1}\mathcal{F}^{1}_{0^{-}}
+ \mathcal{K}_{0^{-}}^{2}\mathcal{F}^{2}_{0^{-}} \right] \,, \\
\mathcal{K}_{0^{-}}^{1}&=&\frac{1}{16\pi^{2}}\int_{0}^{1}\dx{x}K(x)
\mathcal{C}_{02}(\mathfrak{M}^{2}) \,, \nonumber \\
\mathcal{K}_{0^{-}}^{1}&=&-\frac{1}{16\pi^{2}}\frac{2}{M}\int_{0}^{1}\dx{x}
L^{+}(x)\mathcal{C}_{02}(\mathfrak{M}^{2}) \,. \nonumber
\end{eqnarray}

\subsection{Vector channel}

The normalization condition in this channel is:
\begin{multline}
\label{eqn:vnormconst11}
\mathcal{N}^{11}_{1^{--}}=48\int_{0}^{1}\dx{x}
J(x)\left[\left(L^{-}(x)-2J(x)P^{2}\right)
\mathcal{C}_{03}(\mathfrak{M}^{2})\right.
\\ + \left.\mathcal{C}_{02}(\mathfrak{M}^{2})\right] \,.
\end{multline}

\vspace{0.5cm}

\subsubsection{Vector decay constant}

The vector decay constant is given by:
\begin{eqnarray}
\label{eqn:vdecayexpl}
f_{1^{--}}&=&\frac{N_{c}}{3m_{1^{--}}}
\mathcal{K}_{1^{--}}^{1}\mathcal{F}^{1}_{1^{--}} \,,\\
\mathcal{K}_{1^{--}}^{1}&=&\frac{12}{16\pi^{2}}
\int_{0}^{1}\dx{x}\left(L^{-}(x)-2J(x)P^{2}\right)
\mathcal{C}_{02}(\mathfrak{M}^{2}) \,.  \nonumber
\end{eqnarray}

\subsection{Scalar channel}

The normalization condition in this channel is:
\begin{multline}
\label{eqn:snormconst11}
\mathcal{N}^{11}_{0^{++}}=8\int_{0}^{1}\dx{x}J(x)
\left[-\left(L^{+}(x)+2J(x)P^{2}\right)\mathcal{C}_{02}(\mathfrak{M}^{2})\right.
\\ + \left.3\mathcal{C}_{02}(\mathfrak{M}^{2})\right] \,.
\end{multline}

\subsection{Axial vector channel}

\noindent The normalization condition for the axial vector channel
is:
\begin{multline}
\label{eqn:avnormconst11}
\mathcal{N}^{11}_{1^{++}}=48\int_{0}^{1}\dx{x}J(x)
\left[\left(L^{+}(x)+2J(x)P^{2}\right)\mathcal{C}_{03}(\mathfrak{M}^{2})\right.
\\ - \left.\mathcal{C}_{02}(\mathfrak{M}^{2})\right] \,.
\end{multline}

%%% The bibliography %%%

\bibliography{SDEBSEReferences}

\end{document}